\documentstyle[twocolumn,epsf,aps]{revtex}

\begin{document}

\title{\vspace*{-1.0cm}\hfill
	{\normalsize {\tt cond-mat/9505076 }} \vspace*{0.5cm}
	\\
	Renormalization Group Analysis of a
	Noisy Kuramoto-Sivashinsky Equation}

\author{Rodolfo Cuerno and Kent B{\ae}kgaard Lauritsen}

\address{Center for Polymer Studies and Dept.\ of Physics,
	\\
	Boston University, 590 Commonwealth Ave.,
	Boston, Massachusetts 02215~~\cite{email}}

\address{\em (\today)}
\address{~}

\address{
\centering{
\medskip\em
\begin{minipage}{14cm}
{}~~~We have analyzed the Kuramoto-Sivashinsky equation with a stochastic noise
term through a dynamic renormalization group calculation.
For a system in which the lattice spacing is
smaller than the typical wavelength of the linear instability
occurring in the system, the large-distance and long-time behavior of this
equation is the same as for the Kardar-Parisi-Zhang equation
in one and two spatial dimensions.
For the $d=2$ case the agreement is only qualitative.
On the other hand, when coarse-graining on larger scales
the asymptotic flow depends on the initial values of the parameters.
{}~\\
\medskip
{}~
\end{minipage}
}}

\maketitle

%

\narrowtext

\section{Introduction}

Currently, there is much interest in understanding the formation
and roughening of nonequilibrium interfaces
\cite{krug-spohn:1991,family-vicsek:1991,
meakin:1993,barabasi-stanley:1995,hh-zhang:1995}.
A common feature of many interfaces observed experimentally and
in discrete models is that their roughening follows simple scaling laws
characterized by the roughness exponent $\alpha$
and the dynamic exponent $z$,
which determine the scaling behavior of, e.g., the correlation functions.
In many cases, the scaling exponents can be obtained using
stochastic evolution equations
of which a seminal example is the so-called
Kardar-Parisi-Zhang (KPZ) equation \cite{kpz:1986,medina-etal:1989}.

Another equation, which has been
actively discussed in problems of pattern formation, such as
chemical turbulence and flame-front propagation,
is the so-called Kuramoto-Sivashinsky (KS) equation
\cite{ks:1977,sivashinsky:1979,kuramoto:1984,cross-hohenberg:1993}.
This is a deterministic nonlinear equation which
exhibits spatiotemporal chaos.
Qualitatively, the chaotic nature of the KS equation
generates stochasticity in such a way that its solution
displays scaling at large distances and long times.
An important question to answer
is whether the KPZ and KS equations belong to the same or to
different universality classes, i.e., whether the scaling properties
of the interfaces
are described by the same or different values for the critical exponents.
In $1+1$ dimension there are numerical
\cite{zaleski:1989,sneppen-etal:1992,
hayot-etal:1993,li-sander:1995,chow-hwa:1994}
and analytical
\cite{chow-hwa:1994,lvov-etal:1993,yakhot:1981,lvov-procaccia:1992}
results which show that the KS and KPZ equations
indeed exhibit the same scaling behavior.
However,
it is still an open question whether the KS and the KPZ equations fall
into the same universality class in \mbox{$2+1$} dimensions
\cite{yakhot:1981,
lvov-procaccia:1992,procaccia-etal:1992,jayaprakash-etal:1993}.

In order to address these questions from a different point of view,
we have investigated a noisy version of the KS equation
by a dynamic renormalization group (RG) analysis.
Even though the relation of such a noisy KS equation
to the KS system is still to be completely clarified, we believe that
the results we obtain for the scaling behavior of the noisy KS equation
may be suggestive concerning the relation between the KS and KPZ
equations in one and two spatial dimensions.
Moreover, the noisy KS equation studied here may be relevant for the
understanding of dynamic roughening in, e.g., sputter eroded surfaces and,
in principle, to any physical system modeled by the deterministic KS
equation in which the relevance of time dependent noise as, e.g.,
fluctuations in a flux or thermal fluctuations, can be argued for.

The outline of this paper is as follows.
In the next section we introduce the noisy KS equation
and discuss how it naturally arises in the description of ion-sputtering.
Then we derive the RG flow in section \ref{sec:RG-flow}.
Section \ref{sec:results} contains the analysis of the RG flow and our
results for one and two spatial dimensions.
Finally, in section \ref{sec:conclusions} we conclude and summarize.

\section{The Noisy Kuramoto-Sivashinsky Equation}
\label{sec:noisyKS}

In this section we introduce the noisy KS equation and address some
of its peculiarities and limiting cases. We discuss
a physical example in which the noisy KS equation appears
naturally, namely
surfaces eroded by ion-sputtering.

Recently, the dynamics of surfaces undergoing ion-sputtering have been
studied experimentally and several different scaling
behaviors found \cite{eklund-etal,chason,wang}.
Among them it is noteworthy to remark the values
$\alpha = 0.2 - 0.4$, and $z=1.6-1.8$ \cite{eklund-etal},
which are consistent with
the predictions of the KPZ equation in $2+1$ dimensions
\cite{barabasi-stanley:1995,hh-zhang:1995}.
On the other hand,
it is well known that for amorphous targets, ion-sputtering leads in
many cases to the formation of a periodic ripple structure whose
orientation depends on the angle of incidence, $\theta$, of the ions
with the normal to the uneroded surface \cite{carter}.
This periodic structure is associated with an instability in the system.
An experimental study of this kind of morphologies can be found in
\cite{chason}.
In Ref.\ \cite{cuerno-barabasi:1995}
a nonlinear stochastic equation has been proposed to describe the
dynamics of the surface profile height $h(x,t)$ of a two-dimensional
surface sputtered at off-normal angles.
The $\theta \to 0$ limit (normal incidence) of this equation results in
\begin{equation}
        \frac{\partial h}{\partial t}
			= \nu \nabla^2 h -K(\nabla^2)^2 h
			  + \frac{\lambda}{2} (\nabla h)^2 + \eta(x,t)  .
                                                \label{eq:ks}
\end{equation}

Here, we will consider the case
where the field $h(x,t)$ describes the height profile of a $d$-dimensional
surface evolving in a $d+1$-dimensional medium.
The surface tension coefficient $\nu$ is negative, whereas
$K$ is a positive surface diffusion coefficient \cite{herring-mullins}.
The strength of the nonlinearity is given by $\lambda$,
and $\eta(x,t)$ is a Gaussian white noise with zero mean and short-range
correlations described by
\begin{equation}
        \langle \eta(x,t) \eta(x',t') \rangle =2D \delta^d(x-x')\delta(t-t') .
                                        \label{eq:ks-noise}
\end{equation}

\begin{figure}[htb]
\centerline{
        \epsfxsize=7.0cm
        \epsfbox{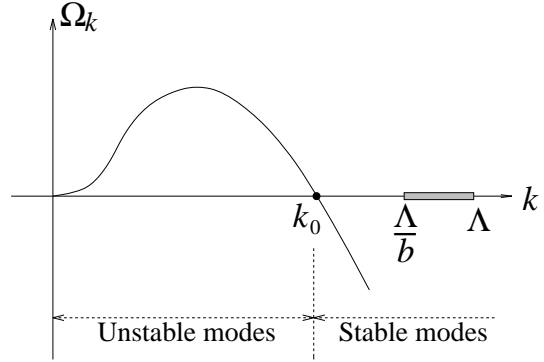}
        \vspace*{0.5cm}
        }
\caption{Stability of Fourier modes for the noisy KS equation.
	The modes from $0$ to $k_0$ are unstable, whereas
	those lying between $k_0$ and $\Lambda$ are stable,
	with $\Lambda$ the momentum cutoff.
	The shaded region corresponds to the momentum shell
	$\Lambda/b < k < \Lambda$ discussed in the text.
        }
\label{fig:modes}
\end{figure}

The fact that $\nu<0$ means that the system is linearly unstable,
a fact which in ion-sputtered systems is related
to faster erosion velocity at the bottom of
the troughs than at the peaks of the crests \cite{sigmund},
which in turn is related to the formation of the periodic ripple structure
referred to above.
The same kind of instability takes place in the deterministic KS
equation, which corresponds to Eq.~(\ref{eq:ks}) without the noise term.
In the following we will refer to Eq.~(\ref{eq:ks}) as the noisy KS equation.
{}From a linear stability analysis
one finds that the amplitude for solutions of the form
\begin{equation}
	h_0(x,t) \sim \exp(\Omega_k t) \sin(kx)
\end{equation}
is characterized by the
rate $\Omega_k = -(\nu k^2 + K k^4)$. This expression is plotted
in Fig.~\ref{fig:modes}, and seen to have a positive value
(corresponding to unstable modes in the system) for
momenta between $0$ and $k_0$, where
\begin{equation}
	k_0 = \sqrt{\frac{|\nu|}{K}}.
					\label{eq:k_0}
\end{equation}
The modes with $k>k_0$ are stable.
In this sense, $k_0$ marks the onset of the unstable modes,
and the length scale $1/k_0$ can be related to the wavelength of the
ripple structure.
The nonlinearity $\lambda$ couples the stable and unstable modes,
thus stabilizing the system \cite{cross-hohenberg:1993}.
If $\nu \ge 0$, it follows that all the modes in the noisy KS equation
are stable and $k_0$ is not defined.

If $\nu$ in Eq.~(\ref{eq:ks})
is a positive coefficient, the contribution of the $K$ term
is expected to be negligible at large length scales and for long times,
so that in this case the noisy KS equation
will show the same scaling behavior as the KPZ equation \cite{kpz:1986}
\begin{equation}
        \frac{\partial h}{\partial t}
                        = \nu \nabla^2 h
                          + \frac{\lambda}{2} (\nabla h)^2 + \eta(x,t) ,
                                                \label{eq:kpz}
\end{equation}
which is obtained as the $K=0$ limit of the noisy KS equation.
However,
if $\nu <0$ in Eq.~(\ref{eq:ks}), as is the case we are interested in,
the surface diffusion
term acts as a stabilizing mechanism at short length scales.
In this situation it is not a priori clear what the scaling
properties of Eq.~(\ref{eq:ks}) are.

\section{RG Flow for the Noisy KS Equation}
\label{sec:RG-flow}

The renormalization group is a standard tool which can be applied to
stochastic equations in order to determine their scaling behavior
in the large-distance long-time hydrodynamic limit \cite{ma:1976}.
Basically, the RG consists of the combination of
a coarse-graining of the system followed by rescaling by a factor $b=e^\ell$.
Successive applications of this transformation leads to
the RG flow of the parameters appearing in the equation in terms of
the scale $\ell$. As usual, we will be interested in considering the
variation of the parameters when $\ell$ is infinitesimal,
that is, the flow is given by first order differential equations
with respect to $\ell$. The critical exponents $\alpha$ and $z$
characterize the scaling behavior of, e.g., the correlation function
\begin{eqnarray}
	C(x-x',t-t') &=& \left< \left[ h(x,t)-h(x',t') \right]^2 \right>
						\nonumber\\
		     &=& |x-x'|^{2\alpha} g \left( |t-t'|/|x-x'|^z \right)  ,
\end{eqnarray}
where $g$ is a scaling function. These exponents are calculated at the
fixed points of the RG transformation.

The standard implementation of the RG procedure to deal with equations
such as Eq.~(\ref{eq:ks}) proceeds first by transforming it to Fourier space.
In our case, it can be easily seen that
the instability ($\nu<0$) in Eq.~(\ref{eq:ks}) generates a pole
for zero frequency at the wavevector $k=k_0$ in the bare propagator
\begin{equation}
	G_0(k,\omega) = \frac{1}{-i\omega + \nu k^2 + K k^4} .
				\label{eq:propagator}
\end{equation}
As a result,
when trying to solve the equation additional divergencies
arise together with those existing for $k \to 0$. The RG circumvents
these divergencies by integrating out a small momentum shell
and can consistently be performed. The momentum shell corresponds
to the wavevectors
\begin{equation}
	\Lambda/b < k < \Lambda  ,
\end{equation}
where the momentum cutoff $\Lambda \equiv 2\pi/a$
is related to the lattice spacing $a$, which acts as a short distance cutoff.
Without loss of generality, we take $\Lambda\equiv 1$ in the following.

The existence of the $k_0$ pole splits the momentum axis in two regions
where the coarse-graining procedure can be applied.
The cutoff $\Lambda$ can either be put in the band of stable modes,
cf.\ Fig.~\ref{fig:modes}, or in the band of unstable modes.
We will discuss the outcome of the RG analysis for both choices.
Intuitively, however, one should coarse-grain over a shell of stable
modes and study the effect of this coarse-graining on the large
scales by means of the RG transformation.

The calculation of the complete RG flow for the noisy KS
equation (\ref{eq:ks}) is tedious. We follow standard diagrammatic
procedures \cite{medina-etal:1989,fns:1977} and here we only give the
final result (see, e.g., \cite{kbl-thesis:1994}).
Compared to the calculations for the KPZ equation ($K=0$)
\cite{kpz:1986,medina-etal:1989,fns:1977}, the main source of the
complications is that we have to keep terms up to order $k^4$ in our
expansion,
in order to be able to calculate the renormalization of the $-K k^4$ term.
The complete RG flow for the noisy KS equation at one-loop order is
\cite{golubovic-bruinsma:1991}
\begin{eqnarray}
        \frac{d\nu}{d\ell} &=& \nu \left( z-2+
				          \frac{\lambda^2 D}{\nu} K_d \,
			     	          \frac{\nu(2-d)+K(4-d)} {4d(\nu+K)^3}
				    \right)  \!,
					              \label{eq:ks-flow-nu} \\
	&&\nonumber\\
        \frac{dK}{d\ell}   &=& \! K \! \left( \! z-4 + \frac{\lambda^2 D}{K}
							K_d \right. \nonumber\\
				&&~~~~~~~~~~~
					\left. \times \frac{a_0\nu^3+a_1\nu^2K
					      +a_2\nu K^2 +a_3 K^3}
				              {16d(d+2)(\nu+K)^5}
					\right) \!\!,\!\!\!\!
				                       \label{eq:ks-flow-K} \\
	&&\nonumber\\
        \frac{d\lambda}{d\ell} &=& \lambda \left( \alpha + z -2  \right) \!,
				                       \label{eq:ks-flow-la} \\
	&&\nonumber\\
        \frac{dD}{d\ell}   &=& D \left( z-2\alpha-d + \lambda^2 D K_d \,
					\frac{1}{4(\nu+K)^3}
			         \right) \!,
				                       \label{eq:ks-flow-D}
\end{eqnarray}
with \mbox{$K_d = S_d/(2\pi)^d$}, where \mbox{$S_d = 2\pi^{d/2}/\Gamma(d/2)$}
is the surface area of the $d$ dimensional unit sphere.
The polynomials $a_i=a_i(d)$ ($i=0,1,2,3$) are given by
\begin{eqnarray}
	a_0 &=& 3d(d-2),~~~~~~~~~~~~~\! a_1 = 11d^2-24d-20,  \nonumber\\
	a_2 &=& 13d^2-40d-60,~~~~ a_3 = 5d^2-22d-16   .
				          \label{eq:a_i}
\end{eqnarray}

When studying the above RG flow,
the coupling constants are conveniently taken to be
\begin{equation}
        g_\nu = \frac{K_d}{4d} \frac{\lambda^2 D}{\nu^3},
		\mbox{~~~~~~}
	f_\nu = \frac{K}{\nu}.
				          \label{eq:gf-nu-def}
\end{equation}
Using these variables,
the flow (\ref{eq:ks-flow-nu})$-$(\ref{eq:ks-flow-D}) reads
\begin{eqnarray}
        \frac{dg_\nu}{d\ell} &=& (2-d) g_\nu
				 + { 4d - 6 + 3(d-4)f_\nu
					\over (1+f_\nu)^3 }  g_{\nu}^{2} ,
	  				             \label{eq:g-nu-flow} \\
	&&\nonumber\\
        \frac{df_\nu}{d\ell} &=& -2f_\nu
  				 + {  b_0 + b_1 f_\nu
					  + b_2 f_{\nu}^{2}
				          + b_3 f_{\nu}^{3}
				          + b_4 f_{\nu}^{4}
  				    \over 4(d+2) {(1+f_\nu)^5} } g_{\nu}.
					                  \label{eq:f-nu-flow}
\end{eqnarray}
The polynomials $b_i=b_i(d)$ ($i=0, 1, \ldots, 4$) are given by
\begin{eqnarray}
	b_0 &=& 3d^2-6d,~~~~~~~~~~~~~~~ b_1=15d^2-24d-36,  \nonumber\\
	b_2 &=& 25d^2-48d-124,~~~~ b_3=17d^2-38d-96, \nonumber\\
	b_4 &=& 4d^2-8d-32    .
				          \label{eq:b_i}
\end{eqnarray}
Note that $g_\nu$ is the only coupling constant one needs to study in
the KPZ case. In our case the flow for $g_\nu$ is affected by the
additional coupling $f_\nu$, which probes the relevance at large
distances of surface diffusion with respect to surface tension.
The pair $(f_\nu, g_\nu)$ is convenient to analyze in the cases where
$K$ is smaller than $\nu$.
On the other hand, when $\nu$ is flowing towards zero the natural
coupling constants are
\begin{eqnarray}
        g_K &=& \frac{K_d}{16d(d+2)} \frac{\lambda^2 D}{K^3}
             =  \frac{1}{4(d+2)} \frac{g_\nu}{f_{\nu}^3},   \nonumber\\
        f_K &=& \frac{\nu}{K} = f_{\nu}^{-1} ,
				          \label{eq:gf-K-def}
\end{eqnarray}
for which the RG flow becomes
\begin{eqnarray}
        \frac{dg_K}{d\ell} &=& (8-d) g_K -
                               { c_0 + c_1 f_K + c_2 f_{K}^{2} + c_3 f_{K}^{3}
                                  \over  (1+f_K)^5  } g_{K}^{2} ,
                                                     \label{eq:g-K-flow} \\
	&&\nonumber\\
        \frac{df_K}{d\ell} &=& 2f_K \! - \!
                                  {b_4 + b_3 f_K + b_2 f_{K}^{2}
                                       + b_1 f_{K}^{3} + b_0 f_{K}^{4}
                                  \over (1+f_K)^5  } g_{K} , \!\!\!
                                                          \label{eq:f-K-flow}
\end{eqnarray}
with the polynomials $c_i=c_i(d)$ ($i = 0,1,2,3$)
\begin{eqnarray}
        c_0 &=& 11d^2-74d-48,~~~~ c_1=31d^2-136d-180,  \nonumber\\
        c_2 &=& 29d^2-80d-60,~~~~ c_3=9d^2-18d  .
				          \label{eq:h_i}
\end{eqnarray}

The flows for Eqs.~(\ref{eq:g-nu-flow})$-$(\ref{eq:f-nu-flow})
and Eqs.~(\ref{eq:g-K-flow})$-$(\ref{eq:f-K-flow})
are singular at the lines $f_{\nu}=-1$ and $f_K=-1$,
respectively. This is the signature of the singularity
(\ref{eq:k_0}) encountered in the bare propagator (\ref{eq:propagator})
for the finite value of the momentum $k=k_0$.
Note also that the two sets of flow equations describe the same system
and, accordingly,
we can analyze the effect of the RG transformation using any of them.

\section{Results and Discussion}
\label{sec:results}

The flow equations (\ref{eq:g-nu-flow})$-$(\ref{eq:f-nu-flow})
and (\ref{eq:g-K-flow})$-$(\ref{eq:f-K-flow}) contain our results for any
dimension $d$.
First, we will determine the fixed points (FP) for the RG flow, and
the values of the critical exponents $\alpha$ and $z$.
Then, we will consider the implications for the flow concerning the large
distance behavior of the system when we coarse-grain in the different
regions of momentum space depicted in Fig.~\ref{fig:modes}.
We will study the physically relevant cases of one and two spatial dimensions.
In Figs.~2$-$5, we have shown the RG flows obtained from the flow equations
for $d=1,2$.
The behavior displayed in the figures is qualitative, and the drawings
are not in scale.
The physical region of the parameter space is determined by
\mbox{$\lambda^2 > 0$}, \mbox{$D>0$}, and \mbox{$K>0$}.
We consider a system which is initially characterized by $\nu<0$,
and in order to determine the renormalization
of the surface tension term we will fix the values of $K$ and $D$.

\begin{figure}[htb]
\centerline{
        \epsfxsize=7.0cm
        \epsfbox{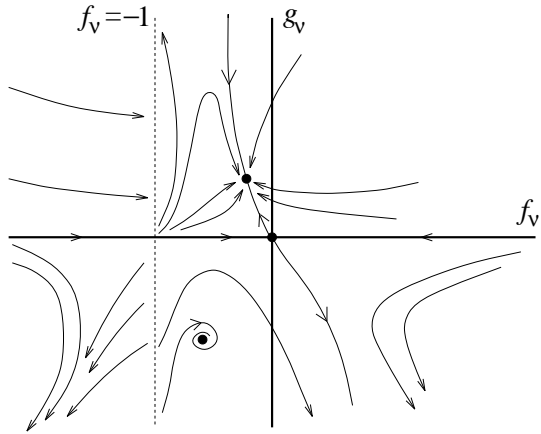}
        \vspace*{0.5cm}
        }
\caption{Schematic RG flow for $(f_{\nu}, g_{\nu})$ in $d=1$.
	There are three fixed points: The EW fixed point
	at the origin, the KPZ attractive fixed point, and a stable spiral FP.
        }
\label{fig:gf-nu-d=1}
\end{figure}

\vspace*{-0.3cm}
\begin{figure}[htb]
\centerline{
        \epsfxsize=7.0cm
        \epsfbox{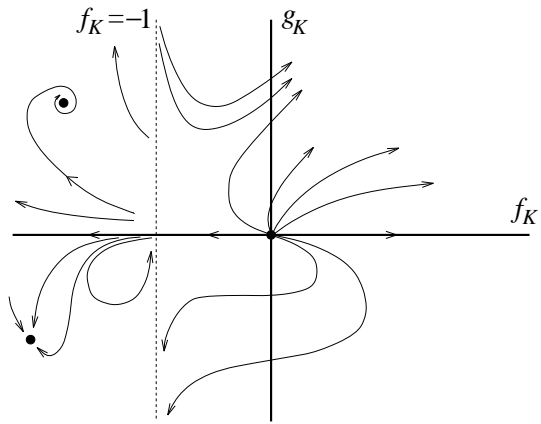}
        \vspace*{0.5cm}
        }
\caption{Schematic RG flow for $(f_{K}, g_{K})$ in $d=1$.
	There are three fixed points: The linear MBE fixed point
	at the origin, the KPZ attractive fixed point, and a stable spiral FP.
        }
\label{fig:gf-K-d=1}
\end{figure}

\subsection{Fixed Points and Critical Exponents}
\label{sec:fixed-points}

As previously stated, the critical exponents are calculated at the
FP's of the RG flow
\mbox{(\ref{eq:ks-flow-nu})$-$(\ref{eq:ks-flow-D})}.
Equation (\ref{eq:ks-flow-la}) reflects the symmetry of
the noisy KS equation under an infinitesimal tilt of the interface
(Galilean transformation)
\cite{kpz:1986,medina-etal:1989,fns:1977} and yields the
exponent identity
\begin{equation}
        \alpha + z = 2    ,
					\label{eq:a+z=2}
\end{equation}
which is valid in any dimension for any finite fixed point
at which $\lambda \neq 0$.

In $d=1$ we find three FP's, cf.\ Fig.~\ref{fig:gf-nu-d=1}.
One is the unstable origin (saddle point), which is characterized by the
Edwards-Wilkinson (EW) exponents \cite{ew:1982}
\begin{equation}
	z=2,~~~~ \alpha=\frac{1}{2} ,
					\label{eq:ew-d=1}
\end{equation}
corresponding to the KPZ equation with $\lambda=0$.
The second one is a stable FP at
$(f_{\nu}^{*}, g_{\nu}^{*}) = (-0.04, 0.539)$
	[or $(f_{K}^{*}, g_{K}^{*}) = (-25.25, -722.8)$]
with exponents
\begin{equation}
        z=1.46,~~~~ \alpha=0.54 .
					\label{eq:kpz-d=1}
\end{equation}
We identify this as the KPZ fixed point and believe the negative value
of $f_{\nu}^{*}$ (i.e., $K^* < 0$) to result from the one-loop approximation.
The third FP is a stable focus (spiral FP, characterized by eigenvalues
of the linearized RG transformation which are imaginary numbers with
negative real parts) at
$(f_{\nu}^{*}, g_{\nu}^{*}) = (-0.248, -1.865)$
	[or $(f_{K}^{*}, g_{K}^{*}) = (-4.037, 10.204)$]
with exponents
\begin{equation}
        z=3.12,~~~~ \alpha=-1.12 .
					\label{eq:3rd-d=1}
\end{equation}

\begin{figure}[htb]
\centerline{
        \epsfxsize=7.0cm
        \epsfbox{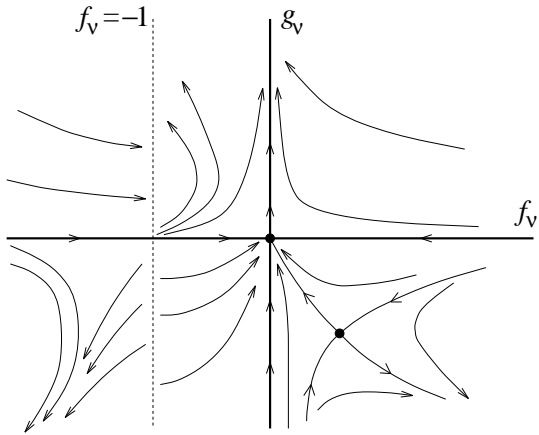}
        \vspace*{0.5cm}
        }
\caption{Schematic RG flow for $(f_{\nu}, g_{\nu})$ in $d=2$.
	There are two fixed points: The EW fixed point
	at the origin and a saddle point. The flow in the
	quadrant $f_{\nu}, g_{\nu} > 0$ is eventually towards the
	KPZ strong-coupling fixed point.
        }
\label{fig:gf-nu-d=2}
\end{figure}

In $d=2$ we find two fixed points, cf.\ Fig.~\ref{fig:gf-nu-d=2}.
One is the unstable origin (saddle point) with the EW
exponents \cite{ew:1982}
\begin{equation}
        z=2,~~~~ \alpha=0 .
					\label{eq:ew-d=2}
\end{equation}
The second one is also an unstable FP (saddle point) at
$(f_{\nu}^{*}, g_{\nu}^{*}) = (1/3, -1.757)$
        [or $(f_{K}^{*}, g_{K}^{*}) = (3, -2.965)$]
with exponents
\begin{equation}
        z=2.49 ,~~~~ \alpha=-0.49 .
					\label{eq:2nd-d=2}
\end{equation}

\begin{figure}[htb]
\centerline{
        \epsfxsize=7.0cm
        \epsfbox{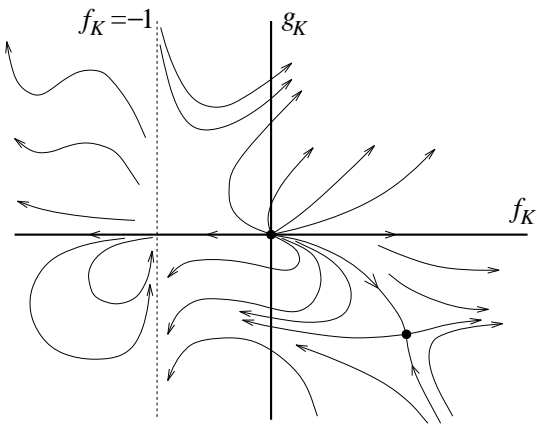}
        \vspace*{0.5cm}
        }
\caption{Schematic RG flow for $(f_{K}, g_{K})$ in $d=2$.
	There are two fixed points: The linear MBE fixed point
	at the origin and a saddle point.
        }
\label{fig:gf-K-d=2}
\end{figure}

Observe that,
in the $(f_{K}^{}, g_{K}^{})$ variables a new FP appears at the origin
for any dimension,
cf.\ Figs.~\ref{fig:gf-K-d=1} and \ref{fig:gf-K-d=2},
with the exponent values
\begin{equation}
        z=4,~~~~ \alpha= \frac{4-d}{2} ,
					\label{eq:mbe}
\end{equation}
corresponding to the linear Molecular-Beam-Epitaxy (MBE) equation
\cite{wolf-villain:1990,villain:1991,lai-dassarma:1991}, which is obtained from
the noisy KS equation (\ref{eq:ks}) by setting $\nu=\lambda=0$.
In our case, this FP is always unstable, which reflects the irrelevance
of surface diffusion at large distances in the presence of surface
tension and a KPZ nonlinearity.

\subsection{RG flow for $\Lambda > k_0$}
\label{subsec:lambda>k_0}

First, let us coarse-grain the noisy KS system in the
band of stable modes, i.e., for $\Lambda > k_0$.
Remember that we use units in which $\Lambda\equiv1$,
so that taking $\Lambda> k_0$ is equivalent to taking $k_0<1$.
In terms of the coupling constants and from Eq.~(\ref{eq:k_0})
this yields $|f_{\nu}|>1$, and since $\nu<0$, this implies
$-\infty < f_{\nu} < -1$.
Therefore, taking $\Lambda > k_0$ implies that we will take initial
values of $\nu$ and $K$ such that
\begin{equation}
	-\infty < f_{\nu} < -1 ,~~~~  g_\nu < 0,
						\label{eq:lambda<k_0}
\end{equation}
or, equivalently,
$$
             -1 < f_K     < 0,~~~~~\, g_K > 0 .
$$

In the remaining subsections we will use the same convention, so that
whenever we discuss the flow for a different value of $\Lambda$
(e.g., $\Lambda \gg k_0$)
this corresponds to taking initial values of $\nu$ and
$K$ such that the value of $k_0$ is in the corresponding relation
to $\Lambda\equiv 1$ (e.g., $k_0 \ll 1$).

\subsubsection{Dimension $d=1$}

We consider the flow (with initially $\nu <0$)
in the $(f_{\nu}, g_{\nu})$ variables, cf.\ Fig.\ \ref{fig:gf-nu-d=1}.
If we take $\Lambda$ close to $k_0$, then $|\nu|$
decreases monotonously, whereas (for a fixed $D$) the quantity
$|\lambda^2/\nu^3|$
increases. This flow takes us to very large values of
$f_{\nu}$ and $g_{\nu}$, where the analysis would eventually
become inconclusive, so that the behavior is more adequately studied in
the $(f_{K}, g_{K})$ variables, cf.\ Fig.~\ref{fig:gf-K-d=1}.
Before doing that, we note that the same
asymptotic behavior is reached if we start out with a value of
\mbox{$\Lambda \gg k_0$}.
In this case, however, initially $|\nu|$ and $\lambda$ grow steadily, until
$\lambda^2$ reaches a value at which $d f_{\nu}/ d \ell = 0$.
Thereafter,
the flow merges with the one already considered for an initial value
$\Lambda \gtrsim k_0$.

Looking at the $(f_{K}, g_{K})$ variables, we find that
starting with  $\Lambda \gtrsim k_0$,
the surface tension increases monotonously (i.e., it becomes less negative),
whereas $\lambda^2$ decreases till it reaches
a line on which $d g_{K}/ d \ell = 0$. {}From that moment on, both $\lambda^2$
and $\nu$ increase, eventually crossing the $\nu=0$ axis ($f_K=0$)
and yielding a flow towards the KPZ fixed point.
If we take $\Lambda \gg k_0$ we reach similar conclusions. First, $|\nu|$
and $\lambda^2$ increase until the flow turns back
on a line along which $d f_{K}/ d \ell = 0$. Thereafter, the flow
merges with that studied for $\Lambda \gtrsim k_0$, and crosses to positive
$\nu$ values. Now, both $f_K$ and $g_K$ grow steadily, so that relevant
information is gained about later stages of the flow if we come back to the
$(f_{\nu}, g_{\nu})$ variables.
In the quadrant $f_{\nu}, g_{\nu} > 0$ for $d=1$
(cf.\ Fig.~\ref{fig:gf-nu-d=1}),
we observe that the flow is towards the fixed point
characterized by the KPZ exponents, cf.\ Eq.~(\ref{eq:kpz-d=1}).

To summarize, the flow which initially started
out with a negative value of $\nu$ has renormalized towards
the KPZ fixed point, which is characterized by a positive value of $\nu$,
when we take the momentum cutoff to lie in the
band of stable modes and coarse-grain the system in that region.
This behavior is in qualitative agreement with the results reported in
Refs.\ \cite{zaleski:1989,hayot-etal:1993,li-sander:1995,chow-hwa:1994},
which investigated the deterministic KS equation by
various coarse-graining procedures different from the method used in the
present paper.
Also, it has been argued in Ref.\ \cite{grinstein:1994}
that for a restricted range of initial parameter values the
noisy KS equation scales to a KPZ equation, which is in accordance with our
results obtained by investigating the complete set of flow equations.

It is instructive to study the evolution under the RG flow of the finite pole
\begin{equation}
	k_0 = \sqrt{\frac{|\nu|}{K}} = \sqrt{|f_K|}  ,
\end{equation}
associated with the instability in the system, cf.\ the discussion
following Eq.~(\ref{eq:k_0}).
A naive scaling argument (which would be correct for the
linear equation) leads to a transformation under rescaling with a
factor $b=e^\ell$ as
\begin{eqnarray}
	\nu & \rightarrow & b^{z-2} \nu ,         \nonumber \\
	K   & \rightarrow & b^{z-4} K ,
\end{eqnarray}
implying that under successive rescalings one would get
	$k_0 \rightarrow b k_0 \rightarrow \infty$,
and we would be left with unstable modes only.
However, we have seen that under the RG flow for the nonlinear equation,
at some point the surface tension $\nu$ becomes zero.
As a result, $k_0 \to 0$,
so that under the RG flow the band of stable modes shrinks to zero,
and the noisy KS system evolves as the stable KPZ equation.

\subsubsection{Dimension $d=2$}

The analysis of the RG flow for $\Lambda > k_0$ in $d=2$ leads to
analogous conclusions to the one-dimensional case.
We study the $(f_{\nu},g_{\nu})$ variables for $-\infty <
f_{\nu} < -1$. The flow in this case is completely analogous to the
$d=1$ case (compare the corresponding regions in
Figs.\ \ref{fig:gf-nu-d=1} and \ref{fig:gf-nu-d=2}),
so that $(f_{\nu},g_{\nu})$ renormalize to very large negative
values. Again, we switch to the $(f_{K},g_{K})$ flow,
cf.\ Fig.~\ref{fig:gf-K-d=2}, where we observe
that $\nu$ is crossing the zero axis, thus renormalizing to a
positive value.
Once we are in the $f_{K},g_{K}>0$ region, the flow is
towards large positive values for $f_{K}$, $g_{K}$, and we return to the
$f_{\nu},g_{\nu} > 0$ variables. Here, we observe that $f_{\nu}$
flows towards the $f_{\nu}=0$ axis (which cannot be crossed), while $g_{\nu}$
steadily increases. This we interpret as the irrelevance of
surface diffusion at large distances, where the flow is eventually
towards the KPZ strong coupling fixed point,
unaccessible to the dynamic RG perturbative approach \cite{hh-zhang:1995}.

\subsection{RG flow for $\Lambda < k_0$}
\label{subsec:lambda<k_0}

We will now analyze the behavior of the noisy KS system
when we integrate out a momentum shell in the band of unstable modes.
The regions in the $(f_{\nu},g_{\nu})$ and $(f_{K},g_{K})$ planes
which correspond to taking a value for the momentum cutoff lying
in the band of unstable modes are
\begin{eqnarray}
	    -1 &<& f_{\nu} < 0 ,~~~~~~~  g_\nu < 0,  \nonumber\\
       -\infty &<& f_{K}   < -1,~~~~    g_K   > 0 ,
						\label{eq:lambda>k_0}
\end{eqnarray}
cf.\ the discussion in the beginning of subsection \ref{subsec:lambda>k_0}.

Taking $\Lambda < k_0$ means that all the modes in the system are
linearly unstable. Consequently,
it would be desirable to check the results
reported in this subsection through some technique of a different nature,
as, e.g., a numerical integration of the noisy KS equation.

\subsubsection{Dimension $d=1$}

In this case, as can be seen in Fig.\ \ref{fig:gf-nu-d=1},
the flow is attracted by the stable focus.
The negative roughness exponent $\alpha$ that characterizes this FP
(cf.\ Eq.~(\ref{eq:3rd-d=1})) could indicate a flat interface.

\subsubsection{Dimension $d=2$}

For a two-dimensional interface, we can see in Fig.\ \ref{fig:gf-nu-d=2}
that the flow starting in the region (\ref{eq:lambda>k_0})
points towards the origin, which as noted above (see Eq.\ (\ref{eq:ew-d=2}))
is characterized by the EW exponents in 2+1 dimensions, i.e.,
the interface is flat.
Interestingly, this is a similar scaling behavior to that
found in Ref.\ \cite{procaccia-etal:1992}
for the scaling solution of the deterministic KS equation in 2+1 dimensions,
in that the same value $z=2$ for the dynamic exponent is also obtained.

\subsection{RG flow for the region $\nu>0$, $K<0$}

For the sake of completeness, we devote the last two subsections
to the analysis of the RG flow in the other parameter regions
of Figs.~\ref{fig:gf-nu-d=1}$-$\ref{fig:gf-K-d=2}.
First, we will consider the case in which the initial
values of the parameters are such that $\nu > 0$ and $K<0$.
In terms of the initial values of the coupling constants,
this corresponds to the region
\begin{eqnarray}
	f_{\nu} &<& 0 ,~~~~   g_{\nu} > 0 ,    \nonumber\\
	f_{K}   &<& 0 ,~~~~   g_{K} < 0 .
\end{eqnarray}
In this case,
the band of stable modes is $[0, k_0]$, whereas the unstable modes
are those with $k>k_0$, that is, the stability of the two momentum
regions has been interchanged with respect to the discussion in
subsections \ref{subsec:lambda>k_0} and \ref{subsec:lambda<k_0}.
Therefore,
if we consider $\Lambda < k_0$, this means that $-1<f_\nu<0$,
and, moreover, that $\Lambda$ lies in the band of stable modes.

In one dimension, when $\Lambda \lesssim k_0$ the flow is towards the singular
line \mbox{$f_{\nu}=-1$}, whereas for
$\Lambda \ll k_0$
the flow can in principle reach the KPZ fixed point,
cf.\ Fig.~\ref{fig:gf-nu-d=1}. The latter
is the behavior one expects for a KPZ equation in which one has
included a negative surface diffusion coefficient, since such a term
would be irrelevant with respect to the stable surface tension term.
Analogous
conclusions can be drawn in the $d=2$ case
(cf.\ Fig.~\ref{fig:gf-nu-d=2}), for which taking
$\Lambda \lesssim k_0$ again leads to a flow towards the line
$f_{\nu}=-1$, whereas for $\Lambda \ll k_0$ the flow is towards
a large positive value of $g_{\nu}$, which we interpret as
pointing towards the strong coupling KPZ behavior.

On the other hand, if we coarse grain a system for which $\Lambda > k_0$,
we place ourselves in the band of unstable modes. Both in one and two
dimensions, it can be seen that $k_0$ flows to a finite value,
\begin{equation}
	k_0 \rightarrow 1 ~~(\equiv \Lambda) ,
\end{equation}
as follows, e.g., from Figs.~\ref{fig:gf-nu-d=1} and \ref{fig:gf-nu-d=2}
in the region where \mbox{$-\infty < f_\nu < -1,~ g_\nu > 0$}.
Consequently,
the singularity remains at a fixed value of the momentum, and
it is reached at some stage along the coarse-graining procedure.
In the flow diagrams, it can be observed that the flow terminates on
the singular line $f_{\nu} = f_K = -1$, both for $d=1,2$,
where the analysis is inconclusive.

\subsection{RG flow for the region $f_{\nu}>0$, $g_{\nu} < 0$}

Finally, we discuss the flow in the region given by
\mbox{$f_{\nu}>0$, $g_{\nu} < 0$}, which
corresponds to the following possibilities for the parameters appearing
in the noisy KS equation:
i)~If both \mbox{$\nu$, $K>0$}, then $\lambda^2 < 0$, or $D<0$,
which is unphysical.
ii)~If both $\nu$, $K<0$, this means we are dealing with a highly
linearly unstable equation.

In $d=1$, fixing $K$ the flow is initially towards small values of $f_{\nu}$
(i.e., $\nu \rightarrow - \infty$), while $\lambda^2$ increases steadily.
The later flow is towards large values of $f_{\nu}$, $g_{\nu}$, so we study
it in the $(f_{K}, g_{K})$ variables.
In terms of these (cf.\ Fig.~\ref{fig:gf-K-d=1}), we
observe that for $f_K \approx 0$, $g_K < 0$ the flow is ultimately towards
the singular line $f_K= -1$, where the analysis is inconclusive.

For $d=2$, there exist two possibilities separated by the saddle FP,
cf.\ Fig.~\ref{fig:gf-nu-d=2}.
If the starting $(f_{\nu}, g_{\nu})$ point lies to
the right of the stable separatrix, the flow is eventually towards the
$f_K = -1$ singular line, following a similar reasoning to that in the
$d=1$ case. However, if the initial value of $(f_{\nu}, g_{\nu})$ is
above the separatrix (as, e.g., starting out with a small $f_{\nu}$,
corresponding to a huge value of $|\nu|$ or to a very small value of $K$),
the flow is towards the origin,
characterized by Edwards-Wilkinson exponents in 2+1 dimensions.

\section{Conclusions}
\label{sec:conclusions}

In the present paper we have introduced a noisy version of the
Kuramoto-Sivashinsky equation, which we have investigated by means of a
dynamic renormalization group procedure.
We have analyzed simultaneously two sets of variables
which are combinations of the parameters appearing in the equation.
For a system in which the lattice spacing is
smaller than the typical wavelength of the linear instability
occurring in the system, the large-distance and long-time behavior of this
equation is the same as for the KPZ equation in one and two spatial
dimensions.
For the $d=2$ case the agreement is only
qualitative, since due to the strong coupling behavior there does not exist
a (finite) fixed point accessible to the dynamic RG perturbative approach.
On the other hand, when coarse-graining on larger scales the asymptotic flow
depends on the initial values of the parameters, and no general
conclusions can be obtained.
It would be interesting to check the results obtained here,
especially those for $\Lambda < k_0$, where all the
modes in the system are unstable,
with the outcome of a qualitatively different approach such as,
e.g., a numerical integration of the noisy Kuramoto-Sivashinsky equation.

Finally, given the fact that the noisy KS equation contains the main
physical mechanisms relevant to the dynamics of sputter-eroded surfaces,
we believe that the results obtained here may be relevant for the
description of the
roughening of such surfaces. Note, however, that the noisy KS equation
is the isotropic limit ($\theta=0$) of the equation derived in
Ref.~\cite{cuerno-barabasi:1995}.
The consequence of this limit for the large distance behavior of the
system is a question that should be elucidated
by the investigation of the full equation
proposed in Ref.~\cite{cuerno-barabasi:1995}, of which the present work
can be thought of as a first step.

\section*{Acknowledgements}

We acknowledge discussions with A.-L.~Barab\'asi and O.~Diego.
Comments and encouragement by H.~Makse, R.~Sadr, and H.~E.~Stanley
are also acknowledged.
R.~C. is supported by a postdoctoral fellowship from the Spanish
Ministerio de Educaci\'on y Ciencia,
and K.~B.~L. is supported by the
Carlsberg Foundation and the Danish Natural Science Research Council.
The Center for Polymer Studies is supported by NSF.


\end{document}